\documentclass[10pt]{iopart}

\usepackage{graphicx}	
\usepackage{epstopdf}
\usepackage{color}
\usepackage{units}
\usepackage{amssymb}
\usepackage[separate-uncertainty=true, multi-part-units = single, product-units=single]{siunitx}
\usepackage{upgreek}

\begin{document}

\title[]{Probing interactions of thermal Sr Rydberg atoms using simultaneous optical and ion detection}

\author{Ryan K. Hanley, Alistair D. Bounds, Paul Huillery, Niamh~C.~Keegan, Riccardo Faoro, Elizabeth M. Bridge, Kevin~J.~Weatherill, Matthew P. A. Jones.}
\address{Joint Quantum Centre (JQC) Durham-Newcastle, Department of Physics, Durham University, South Road, Durham, DH1 3LE, United Kingdom}

\ead{m.p.a.jones@durham.ac.uk}

\begin{abstract}   
We demonstrate a method for probing interaction effects in a thermal beam of strontium atoms using simultaneous measurements of Rydberg EIT and spontaneously created ions or electrons. We present a Doppler-averaged optical Bloch equation model that reproduces the optical signals and allows us to connect the optical coherences and the populations. We use this to determine that the spontaneous ionization process in our system occurs due to collisions between Rydberg and ground state atoms in the EIT regime. We measure the cross section of this process to be $\SI{0.6 \pm 0.2}{}~ \sigma_{\rm{geo}}$, where $\sigma_{\rm{geo}}$ is the geometrical cross section of the Rydberg atom. This results adds complementary insight to a range of recent studies of interacting thermal Rydberg ensembles.

\end{abstract}

\maketitle
\ioptwocol

\section{Introduction}

Rydberg atoms have become a tool of choice to study strongly interacting systems due to their large dipole moments, scaling as $n^2$, where $n$ is the principal quantum number of the atom. These large dipole moments lead to strong Rydberg-Rydberg interactions \cite{Gallagher1994}. In cold atoms, this leads to the so-called ``frozen gas'' regime \cite{Mourachko1998}, where the interaction strength significantly exceeds the kinetic energy, enabling the creation of quantum many-body states \cite{Lukin2001}. Furthermore, using electromagnetically induced transparency (EIT) it has become possible to map these interactions onto light \cite{Pritchard2010} leading to the rapidly growing field of Rydberg quantum optics \cite{Firstenberg2016}.

In thermal vapours, the frozen gas regime does not generally apply, and the interactions have traditionally been viewed as collisional in nature. The collisional properties of Rydberg atoms have been extensively studied using ionization detectors in beam experiments (for a review see \cite{Niemax1985}) for many years. Of particular relevance here are experiments in strontium \cite{Worden1978, Hermann1980, Ye2013}. With the observation of Rydberg EIT in vapour cells \cite{Mohapatra2007} and beams \cite{Mauger2007}, the question of whether interactions can be used to engineer many-body states of atoms and light in thermal vapours has come to the fore \cite{Zhang2015}. A case where interactions do appear to play a decisive role is the observation of Rydberg-mediated optical bistability \cite{Carr2013,Ding2016}, which has lead to the observation of non-equilibrium phase transitions as well as detection methods for terahertz radiation \cite{Wade2016}. However, while theory has shown that an effective mean-field description of the system leads to quantitative agreement with the experiments \cite{Marcuzzi2014,Sibalic2016}, the microscopic nature of this effective interaction remains unclear. In particular, the relative importance of dipole-dipole processes (including e.g. superradiance) and charged particles is an open question \cite{Ding2016,Weller2016}.
 
Since the first observations of Rydberg EIT in a thermal vapour, it has been known that charged particles are produced \cite{Mohapatra2007}. An important step forward was provided by Barredo \textit{et al.} \cite{Barredo2013}, who combined Rydberg EIT with the detection of ions using electrodes placed in a Rb vapour cell. In addition, recent experiments have observed the Stark shift of Rydberg lines due the presence of ions \cite{Weller2016}.	
 	
Here we demonstrate a method for probing interaction effects in a thermal beam of strontium atoms using simultaneous measurements of Rydberg EIT and spontaneously created ions or electrons. Importantly we are able to confirm that our charge detection, based on an electrode that acts as a Faraday cup, is linear for both electrons and ions, despite the presence of secondary charges. By using a model based on the Doppler-averaged optical Bloch equations we are therefore able to quantitatively connect the optical and electrical signals \cite{Lochead2013, Sadler2016, Gavryusev2016}. The results show that spontaneous ionization occurs due to a two-body collision between Rydberg and ground state atoms in the EIT regime.
 
The paper is structured as follows. In section 2, we describe the experimental configuration. Section 3 provides the characterisation of the charge detection system. Section 4 details simultaneous use of both optical and electrical detection methods and section 5 shows the study of the spontaneous ionisation processes in the apparatus.

\section{Experimental Design}\label{fig:appar}
\begin{figure*} 
\includegraphics[width = \textwidth]{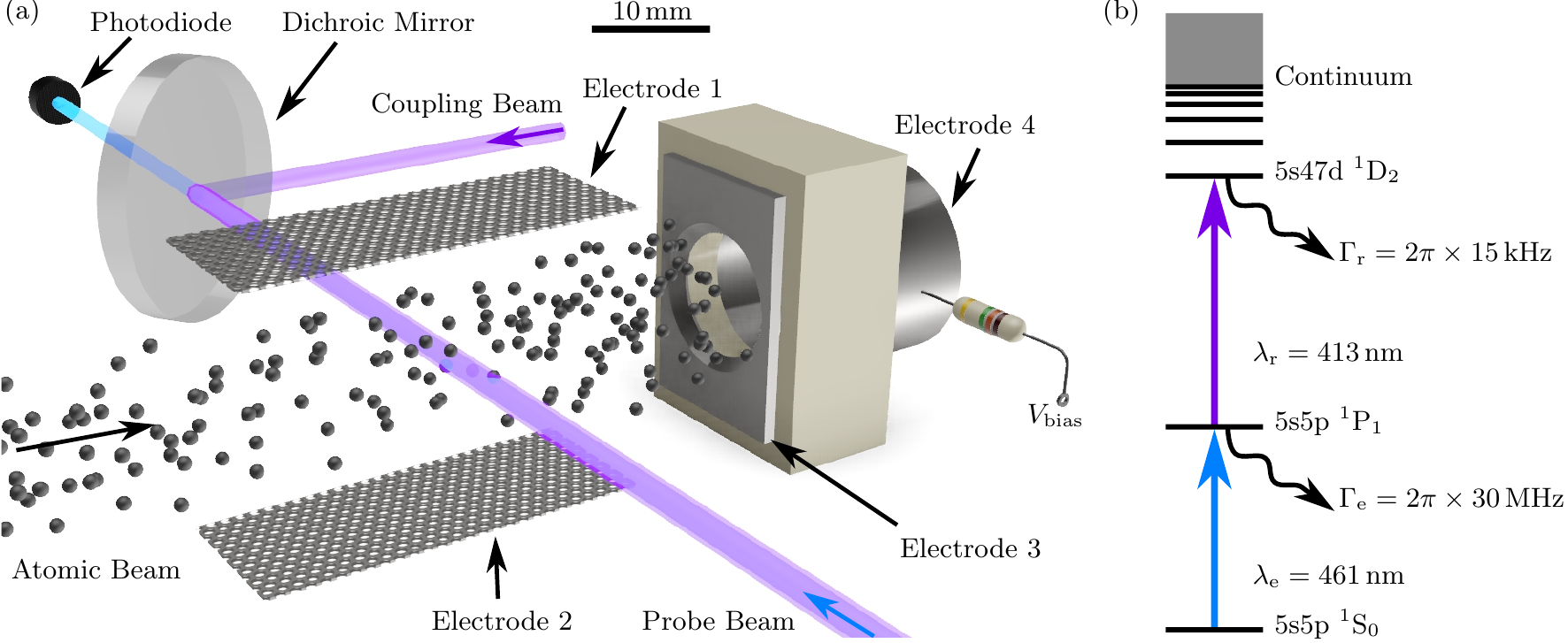} 
\caption{(a) Experimental schematic of the interrogation and detection regions of the beam apparatus. (b) Energy level diagram showing the excitation scheme.}
\label{fig:figure1}
\end{figure*}

We have developed a compact strontium beam apparatus adapted from \cite{Bridge2009} with the addition of electrodes that enable the simultaneous application of electric fields and detection of charged particles. Figure~\ref{fig:figure1}(a) shows the interrogation and detection regions of the beam apparatus. A weakly collimated atomic beam of Sr atoms is produced using a resistively heated dispenser \cite{Bridge2009}. In the interrogation region, Rydberg atoms or ions can be excited using the energy level scheme shown in figure~\ref{fig:figure1}(b), with the laser beams propagating orthogonally to the atomic beam. On the third axis, a pair of stainless steel mesh electrodes (electrodes 1 and 2) allow electric fields to be applied. The mesh design also enables the collection of emitted fluorescence. The separation between the electrodes is  $\SI{21}{mm}$. Beyond the interrogation region, the atomic beam is incident on an additional electrode (electrode~4), which functions as a Faraday cup \cite{Scoles1988}. Charges present in the beam are neutralised at this electrode, causing a current $I$ to flow in an external $\SI{1.3}{M\Omega}$ sense resistor. A further ring electrode (electrode 3) is provided in front of the Faraday cup to provide extra control of the charge detection. 

The probe beam is resonant with the $5{\text{s}}^2~^1{\text{S}}_0 \rightarrow 5{\text{s}}5{\text{p}}~^1{\text{P}}_1$ transition at $\SI{461}{nm}$ and focussed to a $1/e^2$ radius of ${w_x \approx w_y = \SI{115 \pm 1}{\upmu m}}$ at the centre of the interrogation region. The laser frequency is locked using polarisation spectroscopy \cite{Javaux2010} in the same beam apparatus, and the probe beam transmission was measured using a calibrated photodiode. The counterpropagating coupling beam at $\SI{413}{nm}$ is tunable between the ${5{\text{s}}5{\text{p}}~^1{\text{P}}_1 \rightarrow 5{\text{s}}47{\text{d}}~^1{\text{D}}_2}$ transition and the ionisation threshold  \cite{Beigang1982}. It is overlapped with the probe beam using a dichroic mirror, and focussed to  measured waist of ${w_x = w_y = \SI{166 \pm 1}{\upmu m}}$ at the center of the cell. Both laser beams were circularly polarised in order to drive the strongest transitions; $\left|m_j\right| = 0 \rightarrow \left|m_j\right| = 1 \rightarrow \left|m_j\right| = 2$. The coupling beam was amplitude modulated using an optical chopper at a frequency of $\SI{2}{kHz}$. Lock-in detection was thus used to improve the signal-to-noise ratio on both the probe transmission and the current $I$.

\section{Charge Detection}\label{charge}
The Faraday cup (electrode 4) detects charges created in the Sr atomic beam. A bias voltage $V_{\rm{bias}}$ can be applied to the Faraday cup in order to attract or repel positive or negative charges. Since the cup is conductive, charges arriving on the cup are neutralised by a current that flows through the external resistor. The sign of the current depends on the sign of the incident charges. In this paper, we use the convention that positive (negative) current correspond to the detection of negative (positive) charges. In all of the experiments discussed here, the cell body and ring electrode were grounded.

\begin{figure} 
\includegraphics[width = \columnwidth]{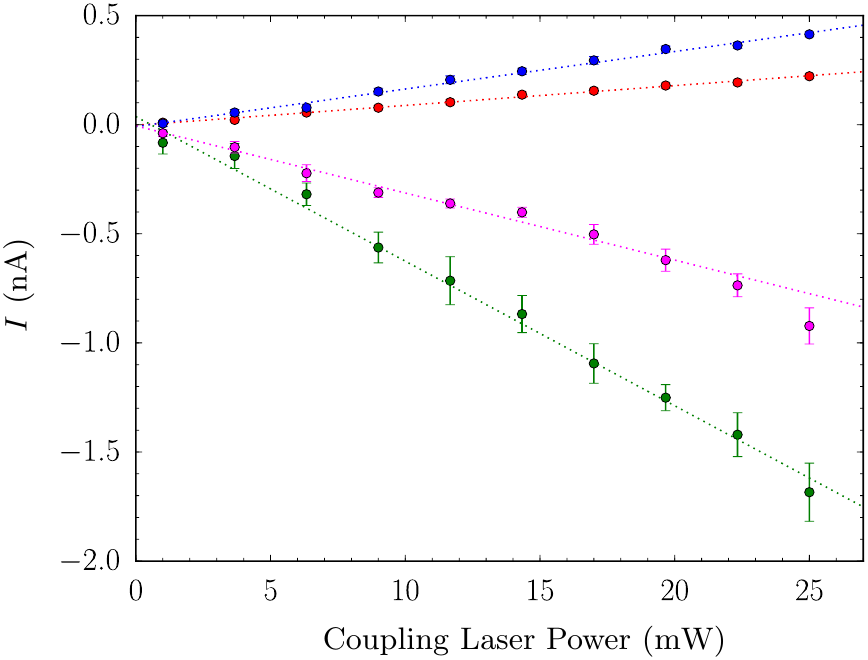} 
\caption{$I$ as a function of coupling laser power. The four curves (top to bottom) correspond to $V_{\rm{bias}} = \SI[retain-explicit-plus]{+5}{V}~ {\rm{(blue)}},~\SI[retain-explicit-plus]{+1.5}{V}~ {\rm{(red)}}~{\rm{and}},~\SI{-1.5}{V}~ {\rm{(purple)}},~\SI{-5}{V}~ {\rm{(green)}}$. The dotted lines are a linear fit to each data set. }
\label{fig:figure2}
\end{figure}

In order to characterise the charge detection system, charges were created in the strontium beam using resonant two-photon photo-ionisation (figure~\ref{fig:figure1}(b)). The coupling laser was tuned $\SI{5}{GHz}$ above the ionisation threshold. In this configuration, the photo-ionisation cross-section is independent of the coupling laser intensity. As long as the probability of photo-ionisation is far from saturated, the ionisation rate is thus directly proportional to the coupling beam power. Therefore, measuring the dependence of the current on the coupling laser beam power provides a test of the linearity of the detection system. Figure~\ref{fig:figure2} shows the measured current $I$ as a function of coupling laser power for a variety of $V_{\rm{bias}}$. A linear fit to the data yields a reduced chi-squared statistic \cite{Hughes2010} of $\chi^2_{\nu}<1.5$ for all values of $V_{\rm{bias}}$, and in each case the fit passes through the origin within the error bar. Our charge detector is therefore linear for currents $\left|I\right|<\SI{2}{nA}$.

\begin{figure}
\includegraphics[width=\columnwidth,keepaspectratio]{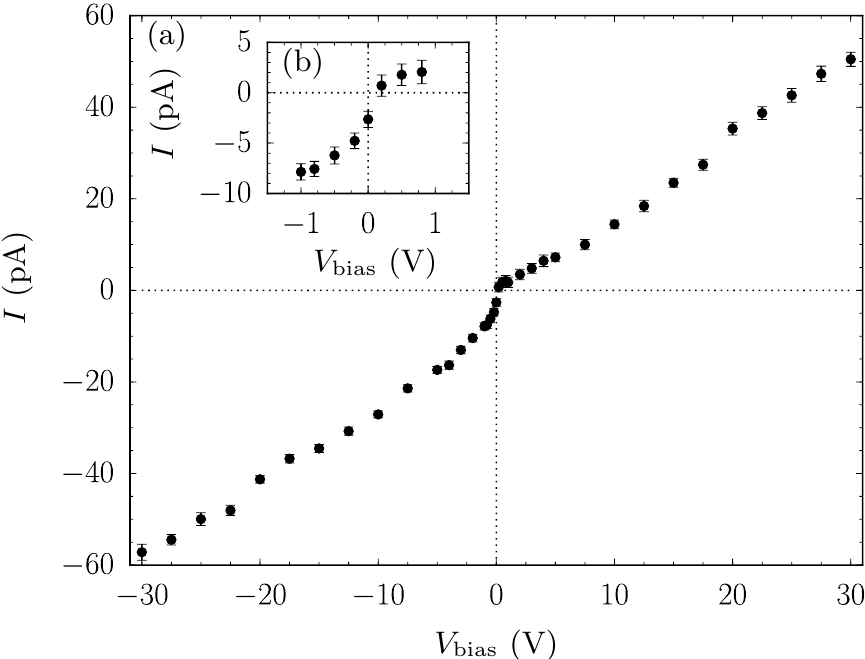}
\caption{(a) Current as a function of $V_{\rm{bias}}$. The probe and coupling powers were $\SI{3.5}{mW}$ and $\SI{33.5}{mW}$ respectively (b) shows a zoom about the origin.}\label{fig:Figure3}
\end{figure}

The effect of varying $V_{\rm{bias}}$ at constant coupling laser power is shown in figure~\ref{fig:Figure3}. There are two clear regions. The first is about the origin, where a small change in  $V_{\rm{bias}}$ results in a large change in current. We attribute this variation to the attraction or repulsion of charges in the atomic beam. The average kinetic energy of the beam is  $\SI{0.1}{eV}$, and consequently only a relatively small voltage is required to modify the trajectory of charges in the beam such that they do or do not hit the detector. The effect of $V_{\rm{bias}}$ will be much greater for electrons than for ions due to their much lower mass. This asymmetry is apparent in figure~\ref{fig:Figure3}(b). A negative current (corresponding to the detection of ions) is visible even for  $V_{\rm{bias}}=\SI{0}{V}$. 

At larger voltages ($|V_{\rm{bias}}|\gtrapprox \SI{1}{V}$) the magnitude of the current increases linearly with the magnitude of $V_{\rm{bias}}$. We attribute this increase to well-known secondary ionisation processes \cite{Lakits1989, Ohya1993}. Incoming charges with energies greater than the work function of the target material can release a charge from the surface which is either re-captured or ejected from the Faraday cup depending on the value of $V_{\rm{bias}}$. 

The Faraday cup detection gain for ions at a particular $V_{\rm{bias}}$ is defined as $G\left(V_{\rm{bias}}\right) = I/I_{\rm{ion}}$. Here $I_{\rm{ion}}$ is the incident ion current in the beam, which can be estimated by using the Doppler-broadened absorption profile on the $5{\text{s}}^2~^1{\text{S}}_0 \rightarrow 5{\text{s}}5{\text{p}}~^1{\text{P}}_1$ transition to infer the atomic flux, and the measured photo-ionization cross-section from \cite{Mende1995, haq2007}. We estimate that $G\left(\SI{0.2}{V}\right)=0.25$.

The linearity of the charge detection over a wide range of current and bias voltages demonstrated in figures \ref{fig:figure2} and \ref{fig:Figure3} is a key requirement for quantitative studies of the spontaneous ionization of thermal Rydberg atoms. As we show in section 4 it enables a direct comparison between experiment and theory, without the  assumptions regarding the excitation and ionization mechanism required in previous work \cite{Barredo2013}.

\section{Simultaneous optical and electronic detection of Rydberg excitation}
\subsection{Rydberg spectroscopy using EIT.}
\begin{figure}
\includegraphics[width=\columnwidth,keepaspectratio]{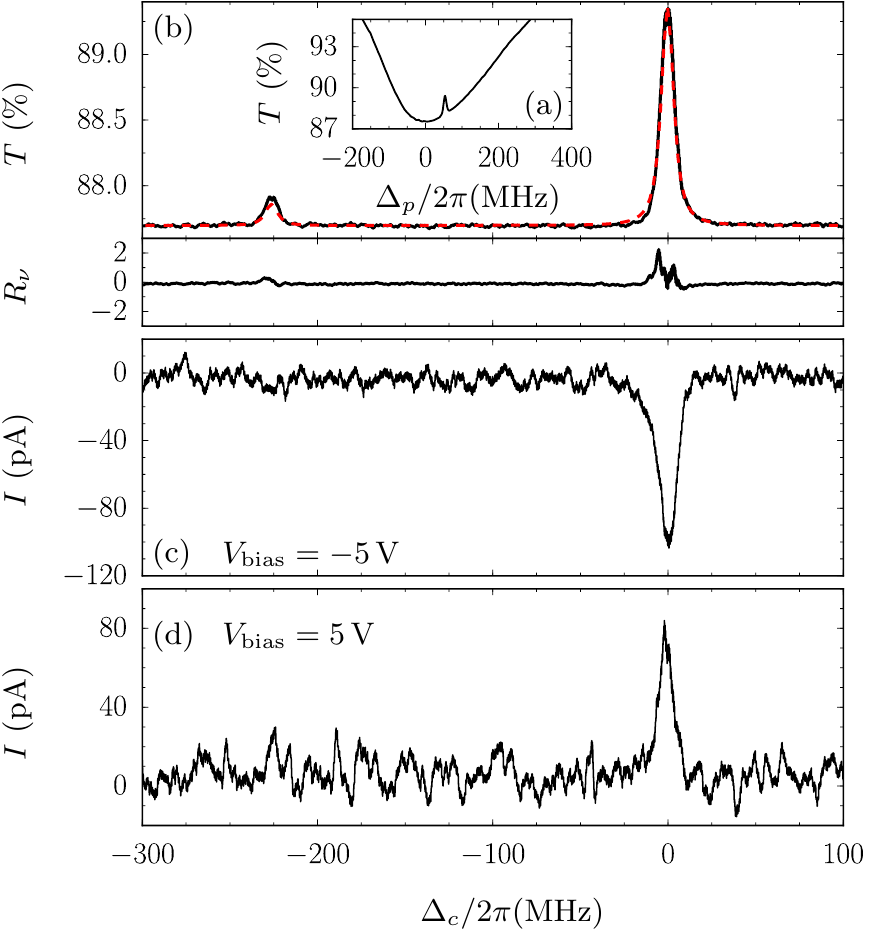}
\caption{(a) Probe transmission as a function of the probe detuning $\Delta _p$ for a fixed coupling beam detuning $\Delta_c$. (b) Probe transmission as a function of coupling beam detuning for fitted coupling Rabi frequency of $\Omega_{\text{c}} /2\pi = \SI{3.1 \pm 0.2}{MHz}$ and coupling laser linewidth $\gamma_{\rm{c}} = \SI{0.6\pm 0.1}{MHz}$. The red dashed line is the theoretical model and the normalised residuals $R_{\nu}$ between the experimental data and the model are shown below. (c) and (d) are the corresponding ion and electron signals.}\label{fig:figure4}
\end{figure}

In order to perform Rydberg spectroscopy in the cell, the coupling laser was tuned close to resonance with the ${5{\text{s}}5{\text{p}}~^1{\text{P}}_1 \rightarrow 5{\text{s}}47{\text{d}}~^1{\text{D}}_2}$ transition. In figure~\ref{fig:figure4}(a), the probe laser frequency was scanned across the ${5{\text{s}}^2~^1{\text{S}}_0 \rightarrow 5{\text{s}}5{\text{p}}~^1{\text{P}}_1}$ transition, and the probe transmission was recorded on an oscilloscope. When the resonance condition for both lasers shown in figure~\ref{fig:figure1} is met, an EIT feature is observed within the Doppler broadened absorption profile. Note that the Doppler-broadened background is not symmetric about zero detuning. This is due to a baffle in the beam apparatus altering the transverse velocity distribution of the strontium beam \cite{Bridge2009}.

A Doppler-free spectrum was obtained by locking the probe laser on resonance, and scanning the coupling laser. The background probe absorption signal can be removed completely by demodulating the probe transmission signal using the lock-in amplifier. The resulting EIT spectrum is shown in figure~\ref{fig:figure4}(b). The largest peak in the EIT signal is attributed to the $^{88}{\rm{Sr}}$ isotope and the smaller peak is attributed to the $^{86}{\rm{Sr}}$ isotope. A peak from the $^{87}{\rm{Sr}}$ isotope is observable at positive detunings but is not shown in this plot. The full width at half maximum (FWHM) of the $^{88}{\rm{Sr}}$ EIT feature is $\SI{11 \pm 1}{MHz}$, which is significantly smaller than the natural linewidth of the probe transition $\Gamma_e$. 
	
The splitting between the EIT features for each isotope provides a measurement of the isotope shift of the ${5{\text{s}}47{\text{d}}~^1{\text{D}}_2}$ state. By calibrating the frequency axis in figure~\ref{fig:figure4}(b) using a high-precision wavemeter, the splitting between the two lines is measured to be ${\Delta \omega_c = 2\pi \times\SI{227 \pm 4}{MHz}}$, where the error quoted is the dominant statistical error. Due to the wavelength mismatch between the probe and coupling lasers, a correction factor must be applied to obtain the isotope shift \cite{Mauger2007}, yielding a value of $\Delta \omega _ 3 = 2\pi \times \SI{213 \pm 4}{MHz}$. This value is slightly smaller than our previous measurement for the ${5{\text{s}}19{\text{d}}~^1{\text{D}}_2}$ state \cite{Mauger2007}, in agreement with the trend observed by Lorenzen \textit{et al.} \cite{Lorenzen1983}.
 
A quantitative model for the EIT lineshape is provided by solving the optical Bloch equations for the three level system \cite{Cohen1998} with broadening due to finite laser linewidths. Here we use the time-dependent solution since the timescale for the evolution of the coherences and the populations is longer than the average time taken for an atom to cross the laser beams ($\approx\SI{0.4}{\upmu s}$). The absorption coefficient is calculated from the coherence between the ground and excited state \cite{Gea1995}. In order to simulate the measured lineshape, the Doppler shift is included in the probe and coupling detunings, and a weighted average is performed over a 1D Boltzmann distribution. The width of the transverse velocity distribution was obtained from the Doppler broadened absorption profile from figure~\ref{fig:figure4}(a). The corresponding effective transverse temperature is $T\approx\SI{60}{K}$. Each isotope was also weighted by its natural abundance. 

The results of the model are shown in figure~\ref{fig:figure4}. The probe laser Rabi frequency ($\Omega_{\text{p}} /2\pi = \SI{13.1}{MHz}$) and linewidth ($\gamma_{\rm{p}} = \SI{0.7 \pm 0.1}{MHz}$) were constrained by independent measurements. The only fit parameters were the coupling Rabi frequency $\Omega_{\text{c}} /2\pi = \SI{3.1 \pm 0.2}{MHz}$ and the coupling laser linewidth $\gamma_{\rm{c}} = \SI{0.6\pm 0.1}{MHz}$. The model is in good agreement with the data, although the residuals show that the theory slightly underestimates the width of the feature. We have checked that this is due to the small amount of additional broadening in the experiment caused by the time response of the lock-in amplifier. 

\subsection{Observation of spontaneous ionisation}\label{ions}
Figure~\ref{fig:figure4} shows that the Rydberg excitation that gives rise to the EIT feature in (a) and (b) also produces a current. Either ions (c) or electrons (d) can be observed simultaneously with the optical signals, using two separate lock-in amplifiers. The signal-to-noise ratio of the current under the conditions of figure~\ref{fig:figure4} is $\geq 4$ (for the $^{88}\text{Sr}$ isotope), allowing us to clearly resolve features and see a direct correspondence between the EIT and electrical signals. We note that the optical and electrical signals have similar widths. In contrast, previous experiments with alkali atoms observed significant broadening of the electrical signal under EIT conditions \cite{Barredo2013}.

\begin{figure}
\includegraphics[width=\columnwidth,keepaspectratio]{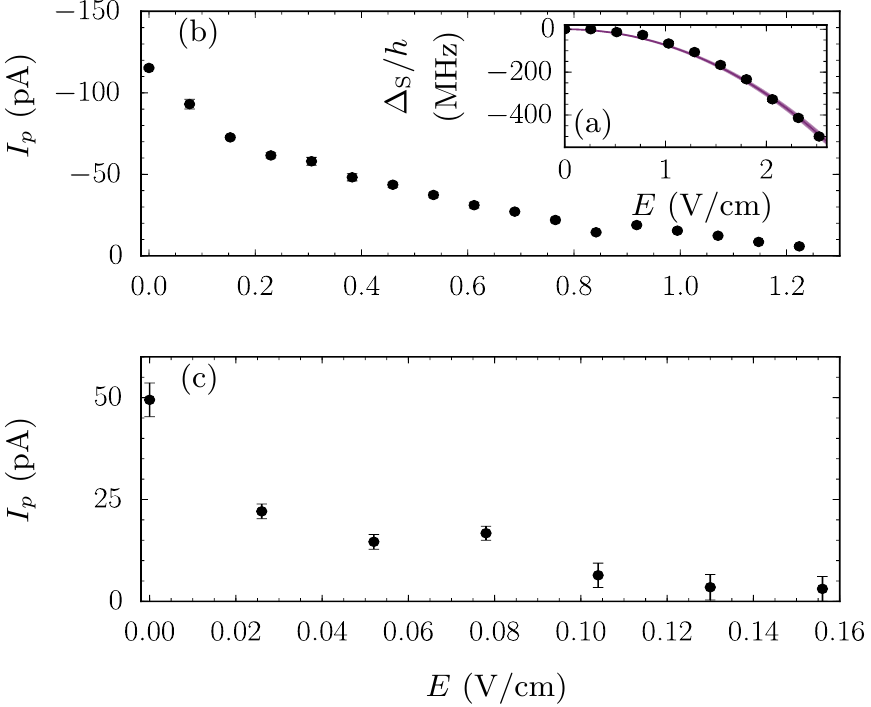}
\caption{(a) Measured Stark shift $\Delta_{\rm{S}}$ of the EIT feature (black circles) along the with theoretical prediction (solid line) as a function of applied electric field $E$. (b-c) $I_p$ as a function of applied electric field for $V_{\rm{bias}} = \SI{-1.5}{V}$ and $V_{\rm{bias}} = \SI[retain-explicit-plus]{1.5}{V}$ respectively.} \label{fig:figure5}
\end{figure}

In order to confirm that the detected charges were not simply created by other mechanisms such as field ionization close to the Faraday cup, we applied a static deflection electric field across the interrogation region. A voltage was applied to electrodes 1 and 2, and the magnitude of the resulting electric field $E$ was calibrated by scanning the coupling laser and measuring the Stark shift of the EIT resonance. For reference, the  electric field required to field ionise the Rydberg atoms is $\approx \SI{66}{V/cm}$ \cite{Gallagher1994}. The resulting Stark map is shown in  \ref{fig:figure5}(a). Due to the laser polarisation, we couple most strongly to the $m_j=0$ state, since the electric field is perpendicular to the propagation direction. The data are compared to a calculated Stark map, which has two constituent parts. Firstly, the expected spatial distribution of the electric field in the interrogation region is calculated using a finite element analysis method (Autodesk Inventor Simulation Mechanical package). Subsequently, the electric field is converted into a predicted line shift by evaluating the Stark shift in the single electron approximation \cite{Millen2011}, using the method of \cite{Zimmerman1979}. Stark maps are calculated by numerical integration of the Stark Hamiltonian, with the necessary quantum defects obtained from \cite{Vaillant2012}. The resulting prediction for the Stark shift is in very good agreement with the data, as shown in figure~\ref{fig:figure5}(a), with no adjustable parameters. The resulting high confidence in the electric field calibration highlights the advantages of this type cell for precision Rydberg spectroscopy \cite{Grimmel2015}. 

The corresponding peak values of the ion (electron) current $I_p$ obtained at the Stark-shifted resonance are shown in figure~\ref{fig:figure5}. As the applied electric field is increased from $\SI{0}{V/cm}$, we see a decrease in the both the electron and ion signal, which is eventually suppressed to the noise floor of our detection system. The electron signal diminishes at much lower fields than the ion signal, as one would expect given the large mass difference of the two particles. These data clearly show that the measured current originates in the interrogation region, and not at the detector. 

\section{Origin of the spontaneous ionisation}
Numerous mechanisms cause Rydberg atoms to spontaneously ionize \cite{Gallagher1994}. They can be usefully classified as one-body or two-body. The former includes collisions with background gas present in the cell and photo-ionization. Photo-ionization may occur due to black-body radiation or the excitation lasers. In two-electron atoms an additional pathway for photo-ionization occurs via off-resonant excitation of the inner valence electron, leading to autoionisation \cite{Gallagher1987}. Two-body ionization can occur due to Rydberg-Rydberg collisions, or due to collisions between Rydberg atoms and another species present in the atomic beam. In our experiments, the possible collision partners are atoms in the ground or intermediate states, as well as ions or electrons. All of these processes have been previously observed in atomic beam experiments \cite{Gallagher1994}.

In this section, we show that we can determine the dominant ionization mechanism by combining the simultaneous measurement of optical and electrical signals with quantitative modelling of the optical response. The model for the optical response also yields a prediction for the population of the Rydberg state, which we compare to the observed current under different assumptions for the ionization mechanism. We note that the observed linearity of the detector (figure \ref{fig:figure2}) is crucial in enabling such a quantitative comparison. Here we only consider the ion signal due to the larger signal-to-noise ratio at lower probe absorptions $A_p$. 

\begin{figure}
\includegraphics[width=\columnwidth,keepaspectratio]{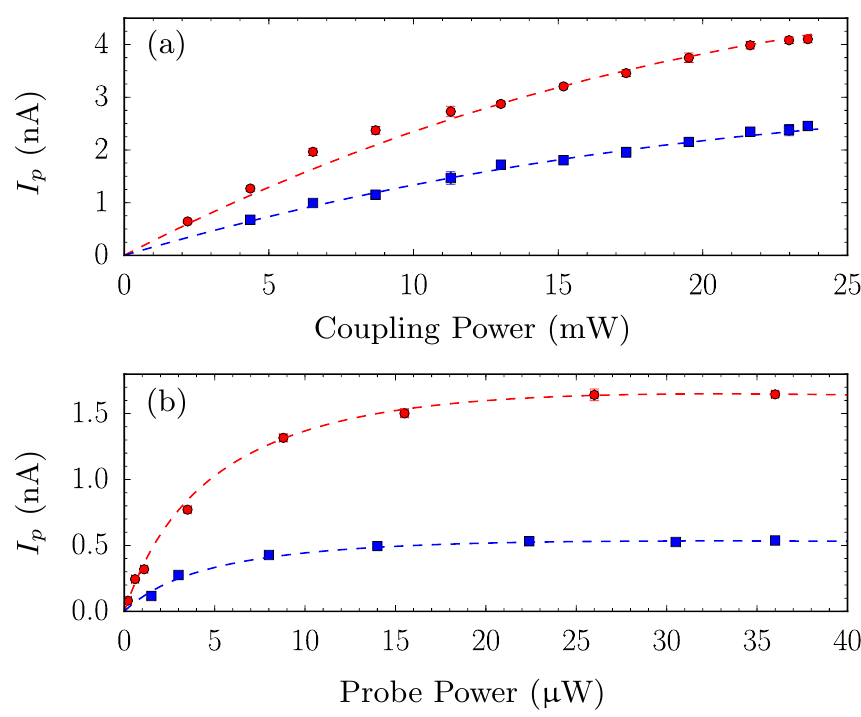}
\caption{Current as a function of coupling beam power (a) and probe beam power (b). The blue (square) and red (circle) points correspond to $\SI{27}{\%}$ and $\SI{32}{\%}$ percentage absorption in the absence of the coupling beam respectively. The dashed lines are the theoretical fit to the data.}\label{fig:figure6}
\end{figure}

Firstly, we varied the population of the Rydberg state at constant ground state density by varying the power of the excitation lasers. The probe and coupling lasers were investigated independently, as shown in figure~\ref{fig:figure6}. Since varying the probe laser also changes the population of the intermediate state while varying the coupling laser does not, a comparison between these experiments is sensitive to any processes that involve the intermediate state.

A prediction for the current is obtained from the optical Bloch model for the optical signal by assuming that the current is related to the Rydberg state population by a combination of one-body and two-body terms: $I_p \propto a_1 N_{\rm{Ryd}} + a_2 N_{\rm{Ryd}}^2$ where $N_{\rm{Ryd}}$ is the number of Rydberg atoms. We find excellent agreement between theory and experiment when the quadratic term is neglected, with $\chi^2_{\nu} \leq 2.2$ for all curves. The implication is that two-body Rydberg-Rydberg processes do not play an observable role. As a result, the current provides a direct measurement of the Rydberg state population. In addition, figure \ref{fig:figure6}(b) suggests that ionization mechanisms involving the intermediate state atoms do not play a role. 

\begin{figure}
\includegraphics[width=\columnwidth,keepaspectratio]{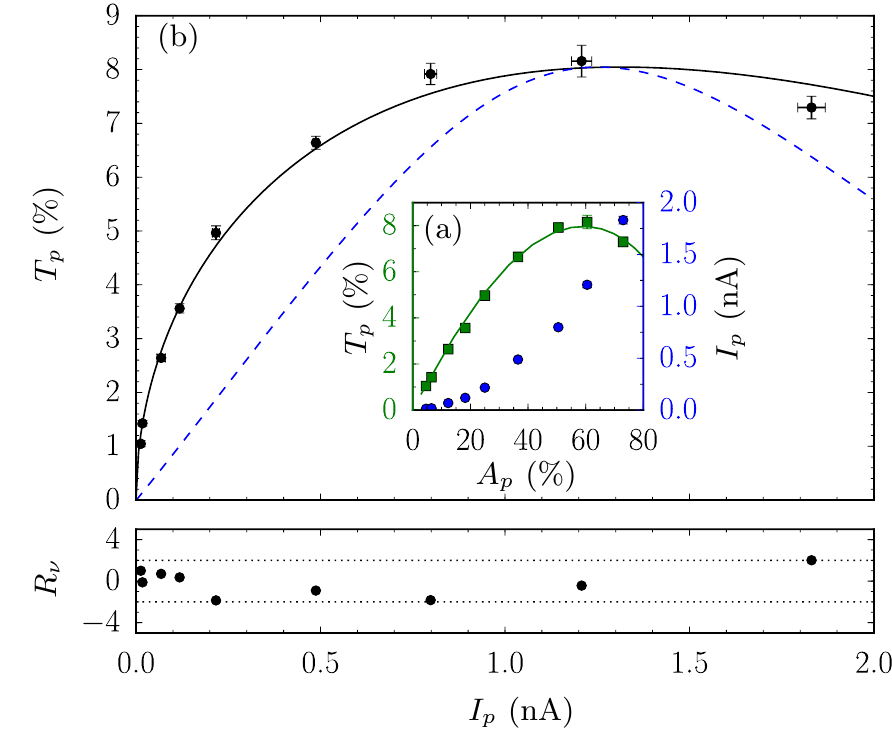}
\caption{(a) $T_p$ (green squares) and $I_p$ (blue circles) as a function of $A_p$. (b) $T_p$ as a function of $I_p$ (black points). The dashed blue curve shows the theoretical model for $I \propto N_{\rm{Ryd}}$. The black curves shows the theoretical model for $I_p \propto a_1 N_{\rm{Ryd}} + a_2 N_{\rm{Ryd}}N_{\rm{Gnd}}$. The residuals between the data and the black curve are normalised to the error bar of each point and are shown below the main figure. The dotted lines show $R_\nu = \pm 2$.}\label{fig:figure7}
\end{figure}

Secondly, we studied the density of the current at constant laser power by varying the flux of the atomic beam. The dependence of the optical response and the current on probe beam absorption is shown in figure \ref{fig:figure7}. At low density, the optical response is proportional to the probe absorption as expected, before exhibiting a roll-off due to propagation effects. At high density absorption of the probe beam leads to a different optical response in different regions of the atomic beam. To take this account in our model, we divide the atomic beam into segments along the probe beam, following the procedure described in \cite{Sadler2016}. The optical Bloch equation model is solved independently in successive segments, with the absorption coefficient in the preceding segment used to set the probe beam intensity in the next. The result is shown in figure ~\ref{fig:figure7}(a) and is in good agreement with the measured probe transmission. In contrast, the current is non-linear in the probe absorption even at low density, strongly indicating that the ionization mechanism is dependent on the ground state density. 

Further insight can be gained by plotting the optical response $T_p$, defined as the difference between the maximum and minimum transmission of the probe beam, as a function of the current $I_p$, as shown in figure~\ref{fig:figure7}(b). The relation between the transmission and the current is clearly non-linear, even at low density, strongly implying that a one-body process (represented by the dashed line) is not responsible the observed ionization. To perform a more quantitative analysis we assume that the Rydberg population calculated by the propagation model is related to the current by a sum of one-body  and two-body (Rydberg-ground state) terms: $I_p \propto b_1 N_{\rm{Ryd}} + b_2 N_{\rm{Ryd}}N_{\rm{Gnd}}$ where $b_1$ and $b_2$ are fit parameters and $N_{\rm{Gnd}}$ is the number of ground state atoms. Here we do not include any two-body terms associated with Rydberg-Rydberg interactions or the intermediate state as these process have previously been excluded (see figure \ref{fig:figure6}). The solid black curve shows a fit with $b_1$ and $b_2$ as adjustable parameters. The best fit is obtained for $b_1=0$, yielding $\chi_{\nu}^2 = 1.9$. This result strongly indicates that the spontaneous ionisation process is not dominated by any of the one-body processes described above. Further emphasis is provided by the lack of agreement between theory and experiment for the dashed blue curve which represents $I_p \propto N_{\rm{Ryd}}$. Furthermore, we note that the absence of a significant one-body ionization rate in addition to a negligible two-body ionization rate in figure \ref{fig:figure6} also eliminates the possibility of collisions between ions and Rydberg atoms

The ionization that we observe is therefore dominated by collisions between ground and Rydberg state atoms, in agreement with \cite{Worden1978}, and the proposals of \cite{Barredo2013,Weller2016}. The most likely cause is Penning ionisation, with the necessary extra energy coming from  the kinetic energy of the atoms \cite{Ben1984}. 

Using a rate equation approach and making a series of approximations, it is possible to estimate the ionisation cross section $\sigma$ of this process. We assume that $\dot{I}_{\rm{Ion}} = \bar{n}\sigma \bar{v}_{\rm{rel}}$ where the dot denotes the time derivative, $\bar{v}_{\rm{rel}}$ is the relative velocity between the ground and Rydberg state atoms \cite{Lubman1982} and $\bar{n}$ is an average density defined as $\bar{n} = \int n_G n_R ~ {\rm{d}}^3r$ where $n_G$ and $n_R$ are the ground and Rydberg state densities respectively. $\sigma = \SI{0.6 \pm 0.2}{}~\sigma_{\rm{geo}}$ where $\sigma_{\rm{geo}} = \pi \left(n^*\right)^4 a_0^2$ is the geometric area of the Rydberg atom and $n^*$ is the effective quantum number \cite{Beigang1982}. The dominant error here is due to the estimation of the interaction volume. To first approximation, one would expect that the cross section is the order of $\sigma_{\rm{geo}}$ \cite{Gallagher1994}, and therefore our value appears reasonable.   

\begin{figure}
\includegraphics[width=\columnwidth,keepaspectratio]{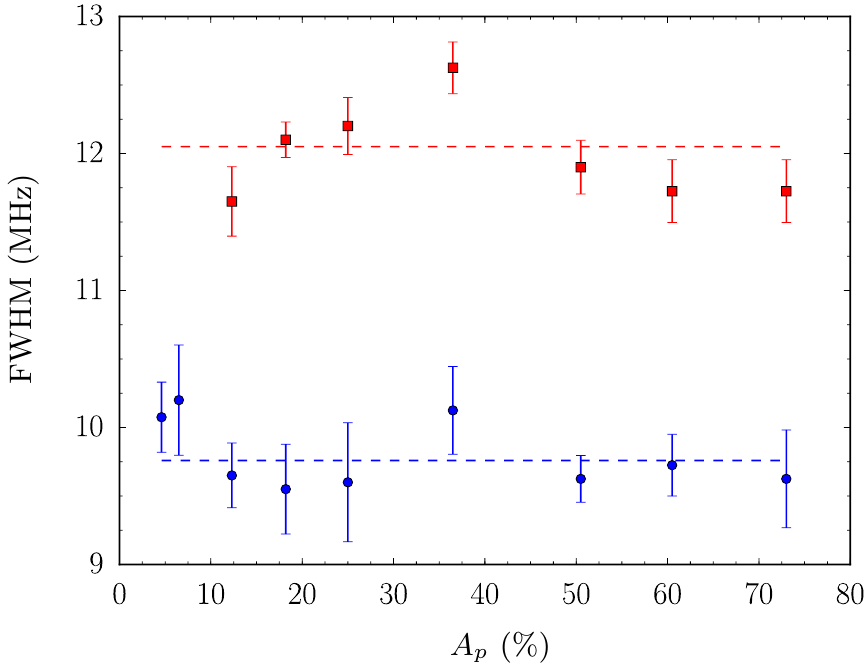}
\caption{Full-width at half-maximum of the EIT feature (blue-circles) and the current feature (red-squares) as a function of $A_p$. The dashed lines are fits to the data where the gradient has been constrained to be zero.}\label{fig:figure8}
\end{figure}

Finally, we return to the evolution of the width of the optical and electrical features with ground state number density. Previous studies have observed significant broadening of the optical and/or electrical response due to pressure broadening \cite{Barredo2013} or interactions \cite{Carr2013,Ding2016,Weller2016}. Figure~\ref{fig:figure8} shows the measured full-width at half-maximum (FWHM) of the EIT and current lineshapes as a function of ground state number density. It is clear that we do not observe any significant broadening over the range of densities we have studied. The measured width of the optical and electrical features are close to those predicted by the optical Bloch equation model, $\SI{8.9}{MHz}$ and $\SI{11.5}{MHz}$ respectively. In particular the model correctly reproduces the difference in width between the two signals. The small amount of extra broadening $\approx\SI{0.7}{MHz}$ that we observe is caused by the time response of the lock-in amplifier. We estimate from the Rydberg density and the measured ion current that the collision rate $\gamma_{\rm{col}} \approx \rm \SI{2}{kHz}$, assuming all collisions result in ion formation. Given that in the absence of collisions, the EIT linewidth is determined by $\Omega_c$ which is much larger than $\gamma_{\rm{col}}$, we would not expect to observe a significant change in the width of the feature due to collisions. 

\section{Concluding Remarks} 
In conclusion, we have performed Rydberg EIT spectroscopy alongside simultaneous measurements of the spontaneous ion signal in a beam apparatus on the ${5{\rm{s}}^2~^1{\rm{S}}_0 \rightarrow 5{\rm{s}}5{\rm{p}}~^1{\rm{P}}_1 \rightarrow 5{\rm{s}}47{\rm{d}}~^1{\rm{D}}_2}$ transitions. The combined analysis of both the optical and electrical signals has shown that the ion signal is a direct measurement of the Rydberg population in our system. It has also been shown to be an effective tool in probing the spontaneous ionisation process, which is dominated by collisions between Rydberg atoms and ground state atoms, where the ionisation energy is derived from the kinetic energy of the atoms. This confirms the hypothesis of the spontaneous ionisation process in previous work \cite{Weller2016,Barredo2013}. This method may provide useful information in other experiments where the ionisation process is currently unknown.

In future work, we will extend this study to the high density regime where Rydberg-Rydberg interactions are significant. The simultaneity of the detection of both optical and electrical signals will shed further light on currently debated topics, such as the role of ions in the formation of Rydberg optical bi-stability \cite{Carr2013,Ding2016,Weller2016}.

\ack
Financial support was provided by EPSRC grant EP/J007021/, Royal Society grant RG140546 and EU grants FP7-ICT-2013-612862-HAIRS, H2020-FETPROACT-
2014-640378-RYSQ and H2020-MSCA-IF-2014-660028. The data presented in this paper are available for download (to be added in proof)

\appendix
\section{}
Here we detail the calculation of the spontaneous ionisation cross section $\sigma$. We assume that ${\dot{I}_{\rm{Ion}} = \bar{n}\sigma \bar{v}_{\rm{rel}}}$. Here the dot denotes the time derivative, $\bar{v}_{\rm{rel}}$ is the relative velocity between the ground and Rydberg state atoms and $\bar{n}$ is an average density. The relative velocity of the atoms in an effusive atomic beam differs from that of a random 3D gas and is given by \cite{Lubman1982} 
\begin{equation}
\bar{v}_{\rm{rel}} = 2\frac{\sqrt{2 k_{\rm{B}} T}}{\pi M} \left(\frac{7 \sqrt{2} - 8}{4}\right)~,
\end{equation}
where $M$ is the mass and $T \approx \SI{800}{K}$ is the temperature of the strontium atoms respectively. The average density is defined as 
\begin{eqnarray}
\bar{n} &= \int n_G n_R ~ {\rm{d}}^3r~, \\
&= n_G n_R A \int {\rm{e}}^{-\Gamma_{\rm{r}} z/\bar{v}}~{\rm{d}}z~, 
\end{eqnarray}
where $n_G$ and $n_R$ are the ground and Rydberg state densities respectively and the integral is performed over the interaction volume. The assumption of the interaction volume is the dominant source of error. Here we assume that the interaction volume is constrained by the the area of the intersection between the probe and atomic beams $A$, and the distance from the probe beam to the Faraday cup.  We integrate over an exponentially decreasing Rydberg distribution due to spontaneous emission. We also assume that the atoms travel with an average velocity $\bar{v}$ towards the Faraday cup. Here, $n_G = \SI{5e14}{m^{-3}}$ is determined from the absorption of the probe beam. Using the optical Bloch equation model, we estimate that the Rydberg population is $n_R = 0.03~n_G$. $A = \SI{6.9E-6}{m^{2}}$ is calculated form the $1/e^2$ diameter of the probe beam and the width of the atomic beam. Using the peak ion signal shown in figure \ref{fig:figure4} and taking into account the gain of the Faraday cup, ${\dot{I}_{\rm{Ion}} = \SI{1e9}{s^{-1}}}$. Using these numbers, we measure the ionisation cross section to be $\sigma = \SI{0.6 \pm 0.2}{}~\sigma_{\rm{geo}}$ where $\sigma_{\rm{geo}} = \pi \left(n^*\right)^4 a_0^2$ is the geometric area of the Rydberg atom and $n^*$ is the effective quantum number \cite{Beigang1982}.

\section*{References}
\bibliographystyle{unsrt}
\bibliography{bibfile}

\end{document}